\documentclass[aps,pre,twocolumn,superscriptaddress,nofootinbib]{revtex4-2}

\usepackage{amsmath}
\usepackage{amssymb}
\usepackage{physics}
\usepackage{indentfirst}
\usepackage{float}
\usepackage{color}
\usepackage{etoolbox}
\usepackage[colorlinks,linkcolor=blue,urlcolor=cyan,citecolor=blue,anchorcolor=blue]{hyperref}
\usepackage[T1]{fontenc}
\usepackage{orcidlink}

\begin{document}
\title{Non-Gaussian transport in tempered fractional Laplace motion}

\author{Yiming Zhao}
\email{Zym916029@mail.ustc.edu.cn}
\affiliation{Department of Physics, University of Science and Technology of China,  Hefei 230026,
China}
\author{Wei Wang\,\orcidlink{0000-0002-1786-3932}}
\affiliation{Institute of Physics \& Astronomy, University of Potsdam, 14476 Potsdam, Germany}
\author{Ralf Metzler\,\orcidlink{0000-0002-6013-7020}}
\email{rmetzler@uni-potsdam.de}
\affiliation{Institute of Physics \& Astronomy, University of Potsdam, 14476 Potsdam, Germany}

\begin{abstract}
We study a minimal stochastic model for active transport featuring a non-Gaussian displacement
probability density function and transient superdiffusion. Our approach is based on a combination
of an exponential tempering in the parental fractional Brownian noise correlations, augmented by
a gamma-distribution of the diffusivity across the ensemble. For the resulting tempered
fractional Laplace motion we derive analytical expressions for the mean-squared displacement,
the kurtosis, the non-Gaussianity parameter and the displacement probability density function.
We also show via stochastic simulations that the model quantitatively reproduces the main
experimental observations in a recent study of active cargo particle transport by a carpet of
amoeba cells.
\end{abstract}

\maketitle

\section{Introduction}

Since Robert Brown first observed the erratic motion of micron-sized granules ejected from pollen
grains when suspended in water \cite{brow1828}, this phenomenon, now known as Brownian motion (BM),
has been detected in numerous thermalised systems and has significantly influenced conceptual
advances in statistical physics \cite{eins1905,smol1906,lang1908,brenig,landau,vankampen}. BM in
one dimension is described in the continuum limit by the diffusion equation \cite{vankampen}
\begin{equation}
\label{diffeq}
\frac{\partial}{\partial t}P(x,t)=K_1\frac{\partial^2}{\partial x^2}P(x,t)
\end{equation}
for the probability density function (PDF) $P(x,t)$, where $K_1$ is the diffusion coefficient of
physical dimension $[K_1]=\mathrm{length}^2/\mathrm{time}$. The analogous equation for a
concentration field is known as Fick's second law \cite{fick,fick1}.

For natural boundary conditions $\lim_{|x|\to\infty}P(x,t)=0$ and sharp initial condition
$\lim_{t\to0}P(x,t)=\delta(x)$ the solution of the diffusion equation (\ref{diffeq}) is the
Gaussian PDF
\begin{equation}
\label{gaussian}
P(x,t)=\frac{1}{\sqrt{4\pi K_1t}}\exp\left(-\frac{x^2}{4K_1t}\right).
\end{equation}
Concurrently, the associated mean-squared displacement (MSD) has the characteristic linear
dependence
\begin{equation}
\label{msd1}
\langle x^2(t)\rangle=2K_1t
\end{equation}
on time. Here, the angular brackets denote ensemble averaging with respect to the PDF $P(x,t)$.

Anomalous diffusion, characterised by the non-linear power-law scaling
\begin{equation}
\label{msd}
\langle x^2(t)\rangle=2K_{\mu}t^{\mu},
\end{equation}
where $\mu\neq1$ is the anomalous diffusion exponent and $K_{\mu}$ of physical dimension $[K_{\mu}]
=\mathrm{length}^2/\mathrm{time}^{\mu}$ denotes the generalised diffusion coefficient, has been
widely observed in a broad range of physical, biological, and geophysical systems
\cite{haus1987diffusion,bouchaud1990anomalous,burov2011single,barkai2012strange,metzler2000report,
metzler2014,igor,brian,brian1}. Anomalous diffusion is non-universal in the sense that physically
different stochastic processes give rise to the same form (\ref{msd}) of the MSD, such that parameter
regression and model classification of anomalous diffusion processes require careful analysis
\cite{yael,henrik,michael,andi,andi1,feng2024reliable,seckler2023machine,qu2024semantic,kevin}.
When $\mu>1$, the motion encoded in the MSD (\ref{msd}) is referred to as superdiffusion, typically
driven by stochastic noise with positive, long-range dependence (long memory), long jump length
statistics, or active motion. Ordinary BM corresponds to normal (Fickian) diffusion with $\mu=1$.
Subdiffusion corresponds to the case $0<\mu<1$, which is often modelled in terms of continuous
time random walks (CTRWs) with scale-free, power-law waiting time PDFs \cite{montroll1965,
mandelbrotwtd} or fractional Brownian motion (FBM) with antipersistent long-range dependence
\cite{kolmogorov,mandelbrot1968}.

Active matter---systems composed of entities that convert internal chemical energy or externally
supplied energy, e.g., via light, into directed motion---has emerged as one of the most rapidly
developing research areas in non-equilibrium statistical physics \cite{bechinger2016active,
marchetti,pawel,sheung2022motor,granek2022anomalous,park2023long,song}. In biological systems,
active transport frequently displays dynamical behaviour that extends far beyond the framework of
classical Brownian motion. Persistent active forces, intermittent interactions, and heterogeneity
of the local environment may act in concert, giving rise to anomalous diffusion and non-Gaussian
displacement statistics \cite{jeon2011,tabei,christine,mend2024,weiss2004,gold2006,christine,
chen2015,beta}. Such signatures are particularly pronounced in composite active-matter systems,
in which passive tracer particles are driven by a surrounding bath of motile agents
\cite{wu2000particle,leptos2009dynamics,mino2011enhanced}.

\begin{figure}
\includegraphics[width=\linewidth]{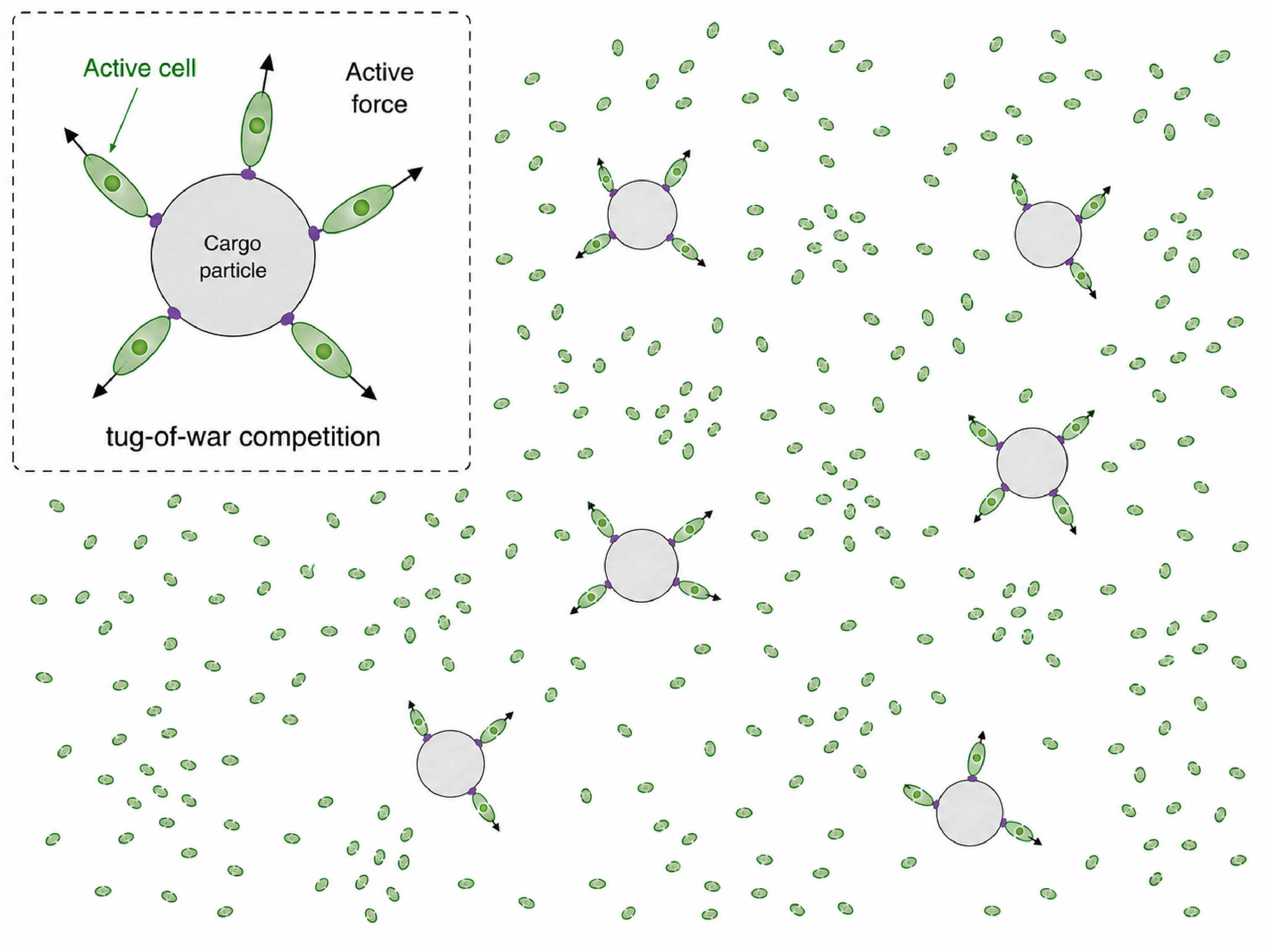}
\caption{Schematic of active transport of passive cargo particles. A quasi-two-dimensional
system contains passive cargo particles and a nonuniform distribution of active cells. While
most cells move freely, a subset adheres directly to cargo particles and exerts active forces.
Inset: A microscopic view of a cargo particle interacting with five active cells, illustrating
the resulting tug-of-war competition between their active pulling exerted on the cargo particle.}
\label{sketch}
\end{figure}

A recent experiment by Gro{\ss}mann \textit{et al.} \cite{grossmann2024non} (see also
\cite{robert}) provides an interesting example for such a composite active system: micron-sized
tracer particles are transported on a carpet of motile \textit{Dictyostelium discoideum\/} cells,
a well-established model organism for amoeboid motion \cite{annesley2009} (see also
\cite{loomis2012,nagel2019,sharifi2024}). Adhesive contacts between the cells and
the tracer particles generate non-thermal fluctuating forces and lead to complex cargo motion,
see the sketch in figure \ref{sketch}. Three characteristic features were reported in this
study \cite{grossmann2024non}: First, the MSD shows a crossover from superdiffusive behaviour
at short lag times to normal diffusion at longer times. Second, the displacement PDF remains
non-Gaussian over the entire experimentally accessible time window, evolving from a strongly
non-Gaussian short-time form toward distributions with pronounced exponential tails. Third, the
non-Gaussianity parameter (see below) is large at short lag times but decreases toward a finite
positive plateau rather than vanishing at long lag times. Thus, while the anomalous scaling of
the MSD is transient, the non-Gaussianity persists.

The long-time non-Gaussian displacement distribution observed in \cite{grossmann2024non} and
similar findings \cite{granick,takumarev,lampo,sabri,kob,sandalo,pastore,vittoria} can be
interpreted in terms of population heterogeneity and a superstatistical distribution of
effective diffusivities \cite{beck2003,beck2006superstatistical,chechkin2017}, stochastically
varying diffusion coefficients ("diffusing-diffusivity" models) \cite{slater,klsebastian,
chechkin2017,indian,vittoriaprl,grebenkov,eiji}, and other approaches \cite{stas,stas1,
luo2018nongaus,hidalgo2020hitchhiker}.

An important concept in the context of random diffusivities is the technique of subordination,
according to which the "operational time" of a stochastic process is replaced by a non-decreasing
random clock process \cite{boch60,feller,foge94,chec21,barkai2001fractional}. As shown in
\cite{chechkin2017} the minimal diffusing-diffusivity model can be rephrased in terms of
subordination.
In particular, subordination by a gamma distribution of diffusion coefficients as known from
financial mathematics \cite{math,math1,math2,kozubowski2006fractional} naturally accounts
for exponential tails of the displacement PDF and a finite non-Gaussianity parameter
\cite{wang2025non}. The description in terms of fractional Laplace motion (FLM), however, does not
capture the full temporal evolution observed in \cite{grossmann2024non}, including the quite
generically expected superdiffusive-to-normal crossover of the MSD along with the strong
non-Gaussianity at long times. Motivated by these observations, we introduce a refined
stochastic model combining three ingredients: (i) FBM with an exponential tempering of the
correlations \cite{molina2018crossover}, which represents the persistent motion of the motile
cells at short times and the gradual loss of directional memory at longer times: transient
superdiffusion followed by normal diffusion; (ii) a L{\'e}vy--Gamma subordination as a
random clock \cite{boch60,foge94,chec21,barkai2001fractional}, accounting for temporal
fluctuations in the effective activity transmitted from the cells to the tracer and generating
strong transient non-Gaussianity; (iii) and a gamma-distribution of generalised diffusivities,
which represents population heterogeneity arising from variations in the cellular environment
and cell--particle coupling and maintains non-Gaussian statistics at long times.

Within this framework, we derive analytical expressions for the MSD, the kurtosis, the
non-Gaussianity parameter, and the (non-Gaussian) displacement PDF. Our analytical predictions
are validated by stochastic simulations. The generalised model reveals two distinct origins of
non-Gaussianity: fluctuations of the operational time dominate at short times but progressively
self-average, whereas the distributed diffusivities sustain a finite non-Gaussian plateau and
exponential displacement PDFs at long times. Consequently, the recovery of normal diffusion in
the MSD does not imply a concurrent crossover to Gaussian displacement statistics. The model
therefore provides a unified description of the transient superdiffusion and persistent
non-Gaussianity observed in active transport on a carpet of motile cells and similar systems.

The remainder of the paper is organised as follows. In section \ref{sec:Model}, we define the
model and explain the physical meaning of our setting. Section \ref{sec:Stat} presents the
statistical properties of the model from theoretical analysis and computer simulations. We
conclude with a discussion and some open problems in section \ref{sec:Conclusion}.

\section{Model and physical observables}
\label{sec:Model}

\subsection{Fractional Laplace motion}
\label{sec:flm-cutoff}

Fractional Laplacian motion (FLM) $x(t)\equiv B_H(s(t))$, is a process resulting from the
subordination of FBM $B_H(s)$ to a L\'evy-gamma process $s(t)$, described by the set of
stochastic equations \cite{kozubowski2006fractional,wang2025non}
\begin{equation}
\label{equ:noise_exp}
\dfrac{dx}{ds}=\sqrt{2D}\xi_H(s),\qquad\dfrac{ds}{dt}=\zeta(t),
\end{equation}
where $s$ is the operational time and $t$ the "laboratory time" of the resulting process.
Here $D$ is the generalised diffusion coefficient of dimension $[D]=\mathrm{length}^2/
\mathrm{time}^{2H}$ and $\zeta(t)$ is the gamma noise. Moreover, $\xi_H(s)$ is fractional
Gaussian noise with zero mean and autocovariance function (ACVF) 
\begin{equation}
\label{fgn}
\langle\xi_H(s_1)\xi_H(s_2)\rangle\sim H(2H-1)|s_1-s_2|^{2H-2},
\end{equation}
whose value can be
positive ("persistent") or negative ("antipersistent"), depending on the magnitude of the
Hurst exponent $H\in(0,1]$. Here the angular brackets denote the noise average. In the
following we restrict our discussion to superdiffusive values, $H\in[1/2,1]$, as will become
clear below.

The subordinator $s(t)$ is the L{\'e}vy-Gamma process characterised by the PDF
\begin{equation}
h(s,t)=\frac{\lambda^{\eta t}}{\Gamma(\eta t)}s^{\eta t-1}e^{-\lambda s},
\end{equation}
where $\lambda$ and $\eta$ are parameters of physical dimension $\mathrm{time}^{-1}$. For the
gamma operational random time $s(t)$, the first two moments can be obtained in the form
\cite{SanchezSilva2016,Bertoin1996}
\begin{eqnarray}
\langle s(t)\rangle=\frac{\eta t}{\lambda},\qquad\langle s^2(t)\rangle=\frac{\eta t(\eta t+1)}{
\lambda^2}.
\end{eqnarray}
In particular, the second moment initially grows linearly in time and subsequently crosses over
to a quadratic time dependence. The coefficient of variation of the process $s(t)$ is then
\begin{eqnarray}
\label{eq-relative}
\frac{\sqrt{\operatorname{Var}[s(t)]}}{\langle s(t)\rangle}=\frac{1}{\sqrt{\eta t}},
\end{eqnarray}
showing that the relative fluctuations of the operational time decay with time, i.e., the random
clock is self-averaging.

FLM belongs to the above-mentioned broader class of subordinated stochastic processes, in which
a parent process evolving in operational time is observed through a random physical-time clock.
This framework has indeed been widely employed to model anomalous transport, heterogeneous and
disordered systems, intermittent dynamics, and non-Gaussian diffusion, as it offers a flexible
means of separating the intrinsic dynamics of the parent process from temporal fluctuations
generated by the random clock \cite{baule2005,eule2009,magdziarz2007fractional,
magdziarz2008equivalence,barkai2001fractional,sokolov2002solutions,chec21}.

In the present context of our guiding example of active cargo transport on a carpet of motile
cells, the random clock provides a phenomenological description of temporal fluctuations in
the effective activity transmitted from the cells to the tracer particle. As cells attach and
detach from the cargo particle, they exert fluctuating forces, such that the intensity of the
active motion may fluctuate strongly over time. The "L{\'e}vy-Gamma clock" captures these
fluctuations by randomly modulating the operational time of the underlying correlated motion,
thereby producing trajectory-to-trajectory variations in the effective amount of activity
experienced over a given physical-time interval. These fluctuations are particularly important
at short times, where they generate pronounced non-Gaussian displacement statistics, whereas
at long times the random clock progressively self-averages and its contribution to the
non-Gaussianity becomes subdominant, as will be discussed in more detail below.

\subsection{Truncated correlation}

For the parental FBM process driven by the fractional Gaussian noise $\xi_H(s)$ with power-law
decay (\ref{fgn}) we now introduce an exponential tempering of the form \cite{molina2018crossover}
\begin{eqnarray}
\nonumber
\langle \xi_H(s_1)\xi_H(s_2)\rangle&=&H(2H-1)|s_1 - s_2|^{2H-2}\\
&&\times\exp\left(-\frac{|s_1-s_2|}{\tau}\right),
\label{eq-fbm-acvf}
\end{eqnarray}
where $\tau$ is the characteristic time for the correlation of the FBM noise. The displacement
PDF of the parental FBM process is Gaussian,
\begin{equation}
G(x,s|D)=\frac{1}{\sqrt{4\pi D\sigma^2(s)}}\exp\left(-\frac{x^2}{4D\sigma^2(s)}\right),
\end{equation}
where we defined the variance
\begin{eqnarray}
\nonumber
\sigma^2(s)&=&\int_0^s(s-\tau)\langle\xi_H(\tau)\xi_H(0)\rangle d\tau\\
\nonumber
&=&H(2H-1)\tau^{2H}\\
&&\times\left[\frac{s}{\tau}\gamma\left(2H-1,\frac{s}{\tau}\right)-\gamma\left(2H,\frac{s}{\tau}
\right)\right],
\label{eq-sigma}
\end{eqnarray}
where $\gamma(a,z)=\int_0^zx^{a-1}e^{-x}dx$ is the lower incomplete gamma function
\cite{olver2010}.

In the present model, the truncated correlation describes the finite persistence of the
cell-driven motion experienced by the tracer. At short times, the power-law correlations of
the parental fractional Gaussian noise retain the directional memory of the active forces
exerted by the motile cells, giving rise to persistent motion and superdiffusive transport.
At times longer than the characteristic correlation time $\tau$, cell reorientation and the
continuously changing configuration of cells coupled to the tracer progressively destroy this
memory. The exponential cutoff therefore limits the duration of the persistent dynamics and
drives the crossover from transient superdiffusion to normal diffusion at long times. In this
sense, $\tau$ sets the characteristic timescale over which the active cellular forcing remains
temporally correlated.

\subsection{Random diffusivity}

On top of our FLM model with exponential tempering of the parental fractional Gaussian noise,
we consider a superstatistical approach \cite{beck2003,beck2006superstatistical} for the
generalised diffusion coefficient $D$. Namely, different trajectories have different diffusion
coefficient distributed according to the gamma distribution
\begin{equation}
\label{equ:f}
f(D)=\frac{1}{\Gamma(\alpha)D_\beta}\left(\frac{D}{D_\beta}\right)^{\alpha-1}\exp(-\frac{D}{D_
\beta}),
\end{equation} 
with the shape parameter $\alpha$ and scaling parameter $D_\beta$. The first and second moments
of the diffusivity are given by
\begin{eqnarray}
\left<D\right>=\alpha D_\beta,\qquad\left<D^2\right>=\alpha(1+\alpha)D^2_\beta.
\end{eqnarray}   

The distributed diffusivity describes the static heterogeneity of the active cellular
environment experienced by different tracer particles, leading to variations in the
effective generalised diffusivity $D$ across the ensemble. We account for this
population heterogeneity by assigning $D$ according to a gamma distribution. In contrast
to the temporal fluctuations generated by the random clock, this "quenched" heterogeneity
does not self-average along an individual trajectory and therefore remains relevant even
at long times. As a result, the ensemble displacement statistics remains non-Gaussian even
after the MSD has crossed over to normal diffusion.

\subsection{Minimal model: tempered fractional Laplace motion}
\label{sec:model-physics}

We now collect the different ingredients described above and establish our minimal model for
active transport on a carpet of motile cells, based on FLM with tempered correlations and
distributed diffusivity. The three ingredients play complementary and physically distinct
roles. The random clock underlying FLM accounts for temporal fluctuations of the effective
cellular activity and generates the pronounced non-Gaussianity observed at short times. The
finite correlation time of the parental process limits the persistence of the cell-driven
motion and produces the crossover from transient superdiffusion to normal diffusion. Finally,
the diffusivity distribution captures static heterogeneity arising from variations in both
the number of cell contacts acting on each cargo particle and the local cell density in its
surrounding environment, which ensures that the displacement statistics remain non-Gaussian even
at long times. Thus, each ingredient is required to reproduce a distinct feature of the
experimentally observed dynamics, making the present construction minimal at the
phenomenological level.

For the present active-transport problem, the random clock $s(t)$ is interpreted as the
accumulated effective activity transmitted from the motile cells to the tracer. Owing to the
stochastic attachment and detachment of cells, fluctuations of the pulling forces, and
changes in the cell--particle coupling, this effective activity does not accumulate uniformly
in physical time. The Gamma clock phenomenologically accounts for these temporal fluctuations.
Its mean is $\langle s(t)\rangle=(\eta/\lambda)t$. In a minimal description, we choose
\begin{eqnarray}
\lambda=\eta,    
\end{eqnarray}
such that $\langle s(t)\rangle=t$. This choice means that, on average, the accumulated
effective activity follows the physical time, while the random clock only introduces
fluctuations around this mean evolution. Thus, it does not introduce an additional systematic
acceleration or slowing down of the parental dynamics \cite{liang}. The remaining clock
parameter $\eta$ controls the magnitude of these temporal activity fluctuations, since
$\mathrm{Var}[s(t)]=t/\eta$, and hence $\eta$ determines the strength and temporal range of
the transient non-Gaussianity. The choice $\lambda=\eta$ removes a redundant time-rescaling
parameter while retaining the essential physical role of the random clock.

Then, the remaining parameters can be associated directly with measurable features of the
dynamics. The Hurst exponent $H$ determines the transient superdiffusive exponent, the cutoff
time $\tau$ controls the crossover from superdiffusion to normal diffusion, the shape parameter
$\alpha$ is fixed by the long-time non-Gaussianity parameter, and the mean generalised
diffusivity $D_\beta$ sets the overall amplitude of the MSD. Consequently, once these
quantities are estimated from the experimental observables, the Gamma-clock parameter
$\eta$ is essentially the only parameter controlling the strength and temporal range of the
transient non-Gaussian fluctuations.

\subsection{Simulations parameters}

The parameters of our minimal model and their exact values used in our computer simulations
are summarised in table \ref{tab:simparams}; for details on the stochastic simulations, see
appendix \ref{app:simulations}. We set the cutoff time $\tau$ to be $10$ time units and
the gamma process rate $\lambda$ to be $10$ inverse time units, such that $\lambda^{-1}\ll
\tau$, which  clearly separates the two timescales in our model. We choose the shape parameter
$\alpha=2$ to be closely aligned with experimental measurements of the diffusivity distribution
\cite{grossmann2024non}.

\begin{table}
\begin{ruledtabular}
\begin{tabular}{lcl}
Parameter & Symbol & Value \\\hline
Hurst exponent & $H$ & $0.7$, $0.9$ \\
Cutoff time (exponential) & $\tau$ & $10$ \\
Gamma process shape & $\eta$ & $10$ \\
Gamma process rate & $\lambda$ & $10$ \\
Diffusivity shape & $\alpha$ & $2$ \\
Diffusivity scale & $ D_\beta$ & $1$ \\
\end{tabular}
\end{ruledtabular}
\caption{\label{tab:simparams}
Parameters for the tempered FLM model with distributed diffusivities used in our simulations.}
\end{table}

\section{Statistical properties}
\label{sec:Stat}

We now analyse the behaviours of the most important statistical observables for our tempered
FLM model.

\subsection{Mean-squared displacement}

Any moment of the tempered FLM process can be obtained by averaging over the parental FBM
process, the gamma subordinator, and the distribution of diffusivities \cite{jakub}
\begin{equation}
\left\langle x^n(t)\right\rangle=\int_{-\infty}^{\infty}x^ndx\int_0^{\infty}dD\int_0^{\infty}ds
G(x,s|D)h(s,t)f(D).
\label{equ:moment}
\end{equation}
Accordingly, the MSD is given by
\begin{eqnarray}
\nonumber
\langle x^2(t)\rangle&=&2\int_0^\infty Df(D)d D\int_0^\infty\sigma^2(s)h(s,t)ds\\
\nonumber
&=& 2H(2H-1)\alpha D_\beta\tau^{2H}\frac{\lambda^{\lambda t}}{\Gamma(\lambda t)}\int_0^\infty
s^{\lambda t-1}e^{-\lambda s}\\
&&\times\left[\frac{s}{\tau}\gamma\left(2H-1,\frac{s}{\tau}\right)-\gamma\left(2H,\frac{s}{\tau}
\right)\right].
\label{eq-msd-gamma}
\end{eqnarray}
Using the integral identity for the incomplete gamma function in Ref.~\cite{olver2010}, 
we obtain (see details in appendix \ref{app-a})
\begin{eqnarray}
\nonumber
\langle x^2(t)\rangle&=&\frac{\alpha D_\beta}{\lambda^{2H}}\frac{\Gamma(\lambda t+2H)}
{\Gamma(\lambda t)}\\
&&\hspace*{-0.8cm}\times{}_2F_1\left(2H-1,\lambda t+2H;2H+1;-\frac{1}{\lambda\tau}\right),
\label{eq-msd}
\end{eqnarray}
where ${}_2F_1$ denotes the Gaussian hypergeometric function \cite{stegun}. We note that the
MSD (\ref{eq-msd}) itself imposes no restriction on the value of $H$. However, the representation
of the MSD in equation (\ref{eq-msd-gamma}) requires $H>1/2$. For the limiting case $H=1/2$, the
hypergeometric function in equation (\ref{eq-msd}) reduces to unity, yielding
\begin{equation}
\label{eq-msd1}
\langle x^2(t)\rangle=\alpha D_\beta\times t,
\end{equation}
which corresponds to normal diffusion.

The asymptotic behaviour of the MSD (\ref{eq-msd}) is governed by two characteristic time
scales: $\lambda^{-1}$, associated with the gamma random clock, and $\tau$, associated with
the tempering of the correlations of the parental FBM process. To clearly separate the
corresponding crossover regimes, we consider the limit $\lambda\tau\gg1$, such that $\lambda^{-1}
\ll\tau$. Combining the three asymptotic regimes, whose detailed derivations are provided in
appendix~\ref{app-a}, we find
\begin{eqnarray}
\hspace*{-0.4cm}
\langle x^2(t)\rangle&\sim&\left\{\begin{array}{ll}\alpha D_\beta\Gamma(2H)\lambda^{1-2H}t,
&t\ll\lambda^{-1},\\[2mm]
\alpha D_\beta t^{2H},&\lambda^{-1}\ll t\ll\tau,\\[2mm]
\alpha D_\beta \Gamma(2H+1)\tau^{2H-1}t,&t\gg\tau.\end{array}\right.
\label{eq:msd_asymptotic}
\end{eqnarray}
The linear short-time behaviour for $t\ll\lambda^{-1}$ can be understood from the strong
fluctuations of the gamma subordinator, where occasional relatively large jumps in the
operational time $s(t)$ contribute significantly to the ensemble-averaged MSD. Since the
probability of such jumps scales linearly with time, the moments of the operational time
satisfy $\langle s^q(t)\rangle\propto t$ for any $q>0$. Consequently, the MSD scales as
$\langle x^2(t)\rangle=2D\langle s^{2H}(t)\rangle\propto t$, independently of the Hurst
exponent $H$ in this limit.

\begin{figure}
(a)\includegraphics[width=0.9\linewidth]{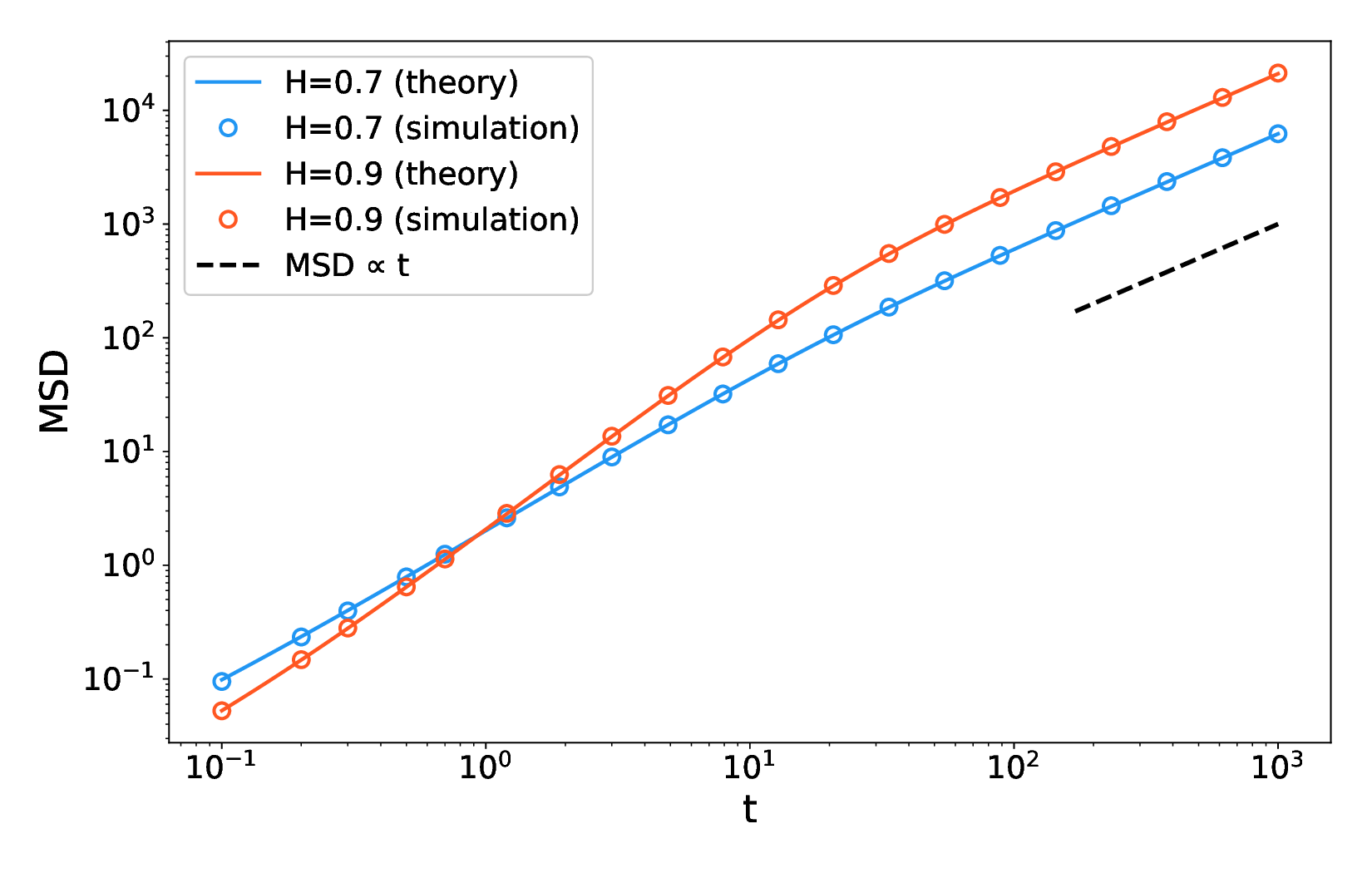}
(b)\includegraphics[width=0.9\linewidth]{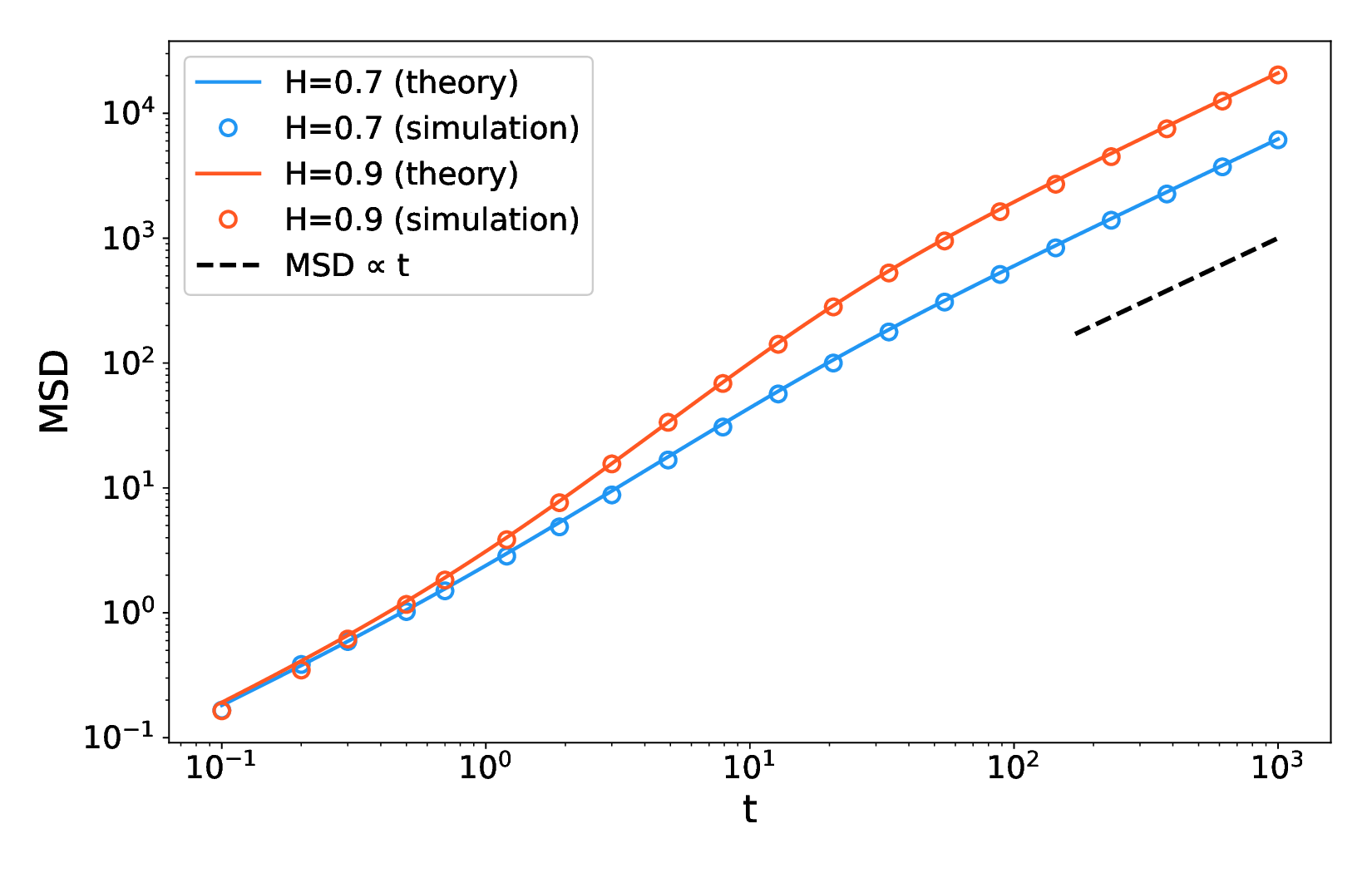}
\caption{(a) MSD of exponentially-tempered FLM with gamma-distributed diffusivity for $H=0.7$
and $0.9$. Solid lines show the theoretical prediction in equation \eqref{eq-msd}, while circles
denote results of stochastic simulations at 20 selected time points. The dashed black line
indicates the linear scaling $\langle x^2(t)\rangle\propto t$ as a guide to the eye. For
details on the simulations see appendix \ref{app:simulations}, the simulations parameters are
listed in table \ref{tab:simparams}. (b) MSD for the case $\lambda=1$, such that the short time
linear scaling can be seen.}
\label{fig:msd}
\end{figure}

The three distinct scaling regimes in equation (\ref{eq:msd_asymptotic}) are each associated with
a different physical mechanism. At very short times, $t\ll\lambda^{-1}$, the MSD grows linearly
with time. This regime originates from the short-time behaviour of the gamma subordinator and is
shifted to increasingly shorter times as $\lambda$ increases. At intermediate times, $\lambda^{
-1}\ll t\ll\tau$, the persistent correlations inherited from the parental FBM dominate, leading to
the anomalous scaling $\langle x^2(t)\rangle\propto t^{2H}$, i.e., superdiffusion for $H>1/2$.
Finally, at long times, $t\gg\tau$, the exponential truncation suppresses the long-range
correlations, and normal diffusion $\langle x^2(t)\rangle\propto t$ is recovered. Thus, the
crossover from the FBM-like superdiffusive regime to the long-time normal-diffusive regime occurs
at the characteristic time scale $\tau$. The different MSD regimes are illustrated in figure
\ref{fig:msd}.

\subsection{Kurtosis and non-Gaussianity parameter}

The kurtosis provides a convenient measure of the non-Gaussianity of the displacement PDF and
is defined as
\begin{equation}
\label{equ:kur_def}
\kappa(t)=\frac{\langle x^4(t)\rangle}{\langle x^2(t)\rangle^2},
\end{equation}
The fourth moment entering equation \eqref{equ:kur_def} can be evaluated from relation
\eqref{equ:moment} following a procedure analogous to that used for the MSD. The resulting
kurtosis is
\begin{subequations}
\begin{equation}
\label{equ:kurtosis}
\kappa(t)=3\frac{\langle D^2\rangle}{\langle D\rangle^2}\frac{\Gamma(\eta t+4H)\Gamma(\eta t)}
{\Gamma^2(\eta t+2H)}\frac{A(t)}{B(t)},
\end{equation}
where
\begin{equation}
A(t)=\int_0^1\int_0^1\frac{(x-1)(y-1)(xy)^{2H-2}}{\left[1+\nu(x+y)\right]^{\eta t+4H}}dxdy,
\end{equation}
and
\begin{equation}
B(t)=\int_0^1\int_0^1\frac{(x-1)(y-1)(xy)^{2H-2}}{\left[(1+\nu x)(1+\nu y)\right]^{\eta t+2H}}dxdy,
\end{equation}
\end{subequations}
with $\nu=1/(\lambda\tau)$.

Equation \eqref{equ:kurtosis} can be naturally decomposed into three contributions. The factor
$\langle D^2\rangle/\langle D\rangle^2$ arises from the distributed diffusivity and quantifies
the population heterogeneity. The ratio of Gamma functions corresponds to the contribution from
the Gamma-subordinated FLM without the correlation cutoff \cite{wang2025non}, while the factor
$A(t)/B(t)$ accounts for the modification introduced by the finite correlation time of the
parental process.

The effect of the finite correlation time on the kurtosis is entirely contained in the ratio
$A(t)/B(t)$. As shown in appendix \ref{app-b}, this ratio is bounded according to
\begin{equation}
\left(1+\frac{1}{\lambda\tau}\right)^{-2H}<\frac{A(t)}{B(t)}<1.
\label{equ:bound}
\end{equation}
Therefore, in the regime $\lambda\tau\gg1$, one has $A(t)/B(t)=1+\mathcal{O}\!\left[(\lambda\tau)^{
-1}\right]$, where $\mathcal{O}$ denotes the Landau symbol. The kurtosis can then be approximated
as
\begin{equation}
\kappa(t)=3\left(1+\frac{1}{\alpha}\right)\frac{\Gamma(4H+\lambda t)\Gamma(\lambda t)}{\Gamma^2
(2H+\lambda t)}\left[1+\mathcal{O}\left(\frac{1}{\lambda\tau}\right)\right].
\end{equation}
This suggests that the timescale $\tau$ in the FBM correlation cutoff does not influence the
kurtosis, which appears reasonable as the cutoff in the FBM correlation function does not change
its Gaussian character. 

In the short-time limit $\lambda t\ll1$, using $\Gamma(\lambda t)\sim1/(\lambda t)$, the kurtosis
diverges as
\begin{equation}
\kappa(t)\sim3\left(1+\frac{1}{\alpha}\right)\frac{\Gamma(4H)}{\Gamma^2(2H)}\frac{1}{\lambda t}.
\end{equation}
This divergence is a characteristic consequence of the Gamma subordination: at times shorter than
the characteristic time scale of the Gamma process, the operational time $s$ exhibits strong
trajectory-to-trajectory fluctuations, leading to pronounced variations in the particle
displacement, see also the discussion in \cite{wang2025non}.

In the long-time limit $\lambda t\gg1$, the Gamma-function ratio approaches unity, and the
kurtosis converges to
\begin{equation}
\label{equ:kurtosis_long}
\kappa(t)\sim3\left(1+\frac{1}{\alpha}\right).
\end{equation}
For finite $\alpha$, this limiting value remains larger than the Gaussian value $\kappa=3$.
Hence, although the transient contribution generated by the Gamma random clock progressively
self-averages, the displacement PDF remains non-Gaussian at long times due to the persistent
diffusivity heterogeneity.

\begin{figure}
(a)\includegraphics[width=0.9\linewidth]{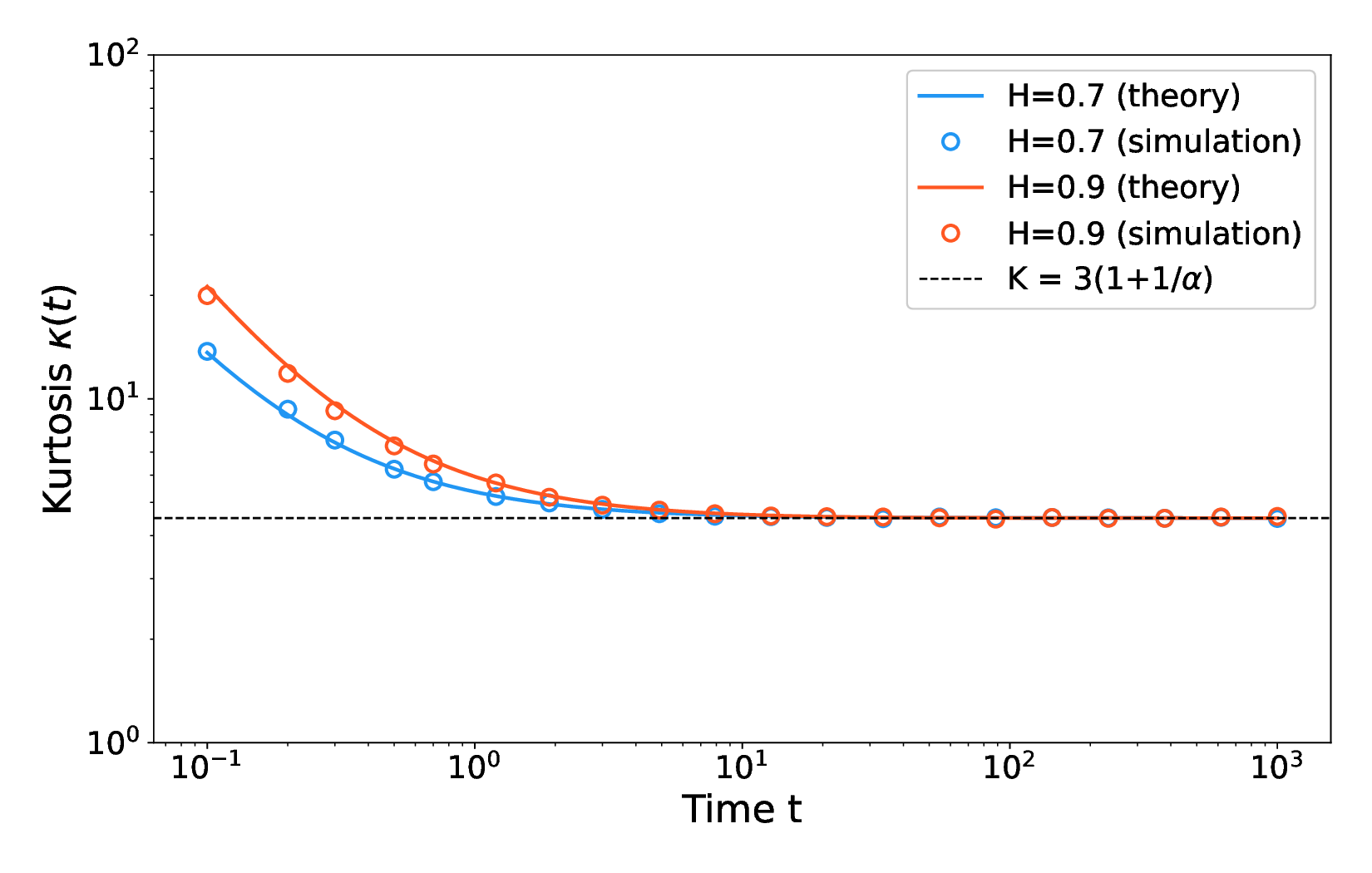}
(b)\includegraphics[width=0.9\linewidth]{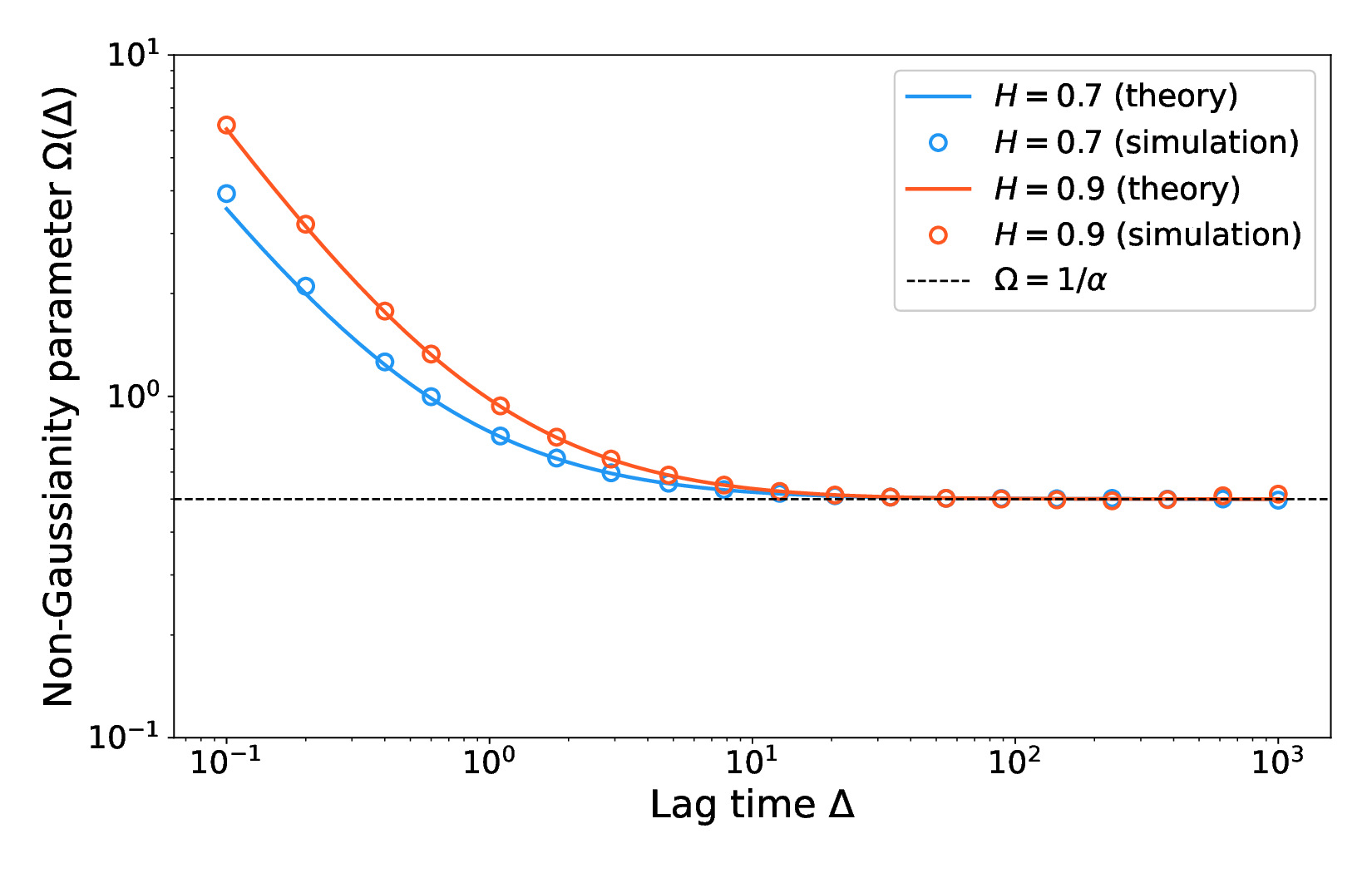}
\caption{Kurtosis (a) and non-Gaussianity parameter (b) for exponentially-tempered FLM with
gamma-distributed diffusivity for Hurst index $H=0.7$ and $0.9$. (a) Theoretical result (solid
curve) from equation \eqref{equ:kurtosis} and simulation result (circles) for the kurtosis. The
black dashed line is the kurtosis at long times as predicted by equation \eqref{equ:kurtosis_long}.
(b) Simulations results for the non-Gaussianity parameter. The black dashed line corresponds to
$y=1/\alpha$ when $\alpha=2$.}
\label{fig:ngp}
\end{figure}
 
Gro{\ss}mann \textit{et al.} \cite{grossmann2024non} considered the closely related measure
given by the non-Gaussianity parameter. Here we consider its one-dimensional version
\cite{rahman1964correlations},
\begin{equation}
\label{eq-ngp}
\Omega(\Delta)=\dfrac{1}{3}\dfrac{\expval{\overline{\delta^4(\Delta)}}}{\expval{\overline{
\delta^2(\Delta)}}^2}-1,
\end{equation}
where $\overline{\delta^n(\Delta)}$ denotes the time-averaged $n$th-order displacement moment
of a single trajectory $x(t)$ of total measurement time $T$ \cite{burov2011single,
barkai2012strange},
\begin{equation}
\overline{\delta^n(\Delta)}=\frac{1}{T-\Delta}\int_0^{T-\Delta}|x(t+\Delta)-x(t)|^ndt.
\end{equation}

We first note that the present process has stationary increments. The parental process $B_H(s)$
is driven by stationary fractional Gaussian noise whose autocovariance depends only on the time
difference, as evident from equation (\ref{eq-fbm-acvf}), and therefore $B_H(s+\Delta s)-B_H(s)$
depends in distribution only on $\Delta s$. Moreover, the Gamma subordinator $s(t)$ is a L{\'e}vy
process with stationary independent increments, such that $s(t+\Delta)-s(t)$ depends only on the
physical lag $\Delta$. Since the parental process and the random clock are independent, the
subordinated process $x(t)=B_H[s(t)]$ consequently satisfies
\begin{equation}
x(t+\Delta)-x(t)\overset{d}{=}x(\Delta)-x(0),
\end{equation}
where $\overset{d}{=}$ denotes equality in distribution. The additional averaging over the
trajectory-dependent diffusivity $D$ does not alter this property. Hence,
\begin{equation}
\label{eq-stationary}
\left<\overline{\delta^n(\Delta)}\right>=\left<|x(\Delta)-x(0)|^n\right>.
\end{equation}

For the one-dimensional non-Gaussianity parameter (\ref{eq-ngp}), the stationary property of
the increment process $x(t+\Delta)-x(t)$, as shown in (\ref{eq-stationary}), immediately
leads to
\begin{equation}
\label{eq:ngp_kurtosis}
\Omega(\Delta)=\frac{1}{3}\frac{\langle x^4(\Delta)\rangle}{\langle x^2(\Delta)\rangle^2}-1=
\frac{\kappa(\Delta)}{3}-1.
\end{equation}
Figure \ref{fig:ngp} shows the simulation results for the kurtosis and the non-Gaussianity
parameter, both of which are in excellent agreement with the theoretical predictions.

\begin{figure*}
(a)\includegraphics[width=7.2cm]{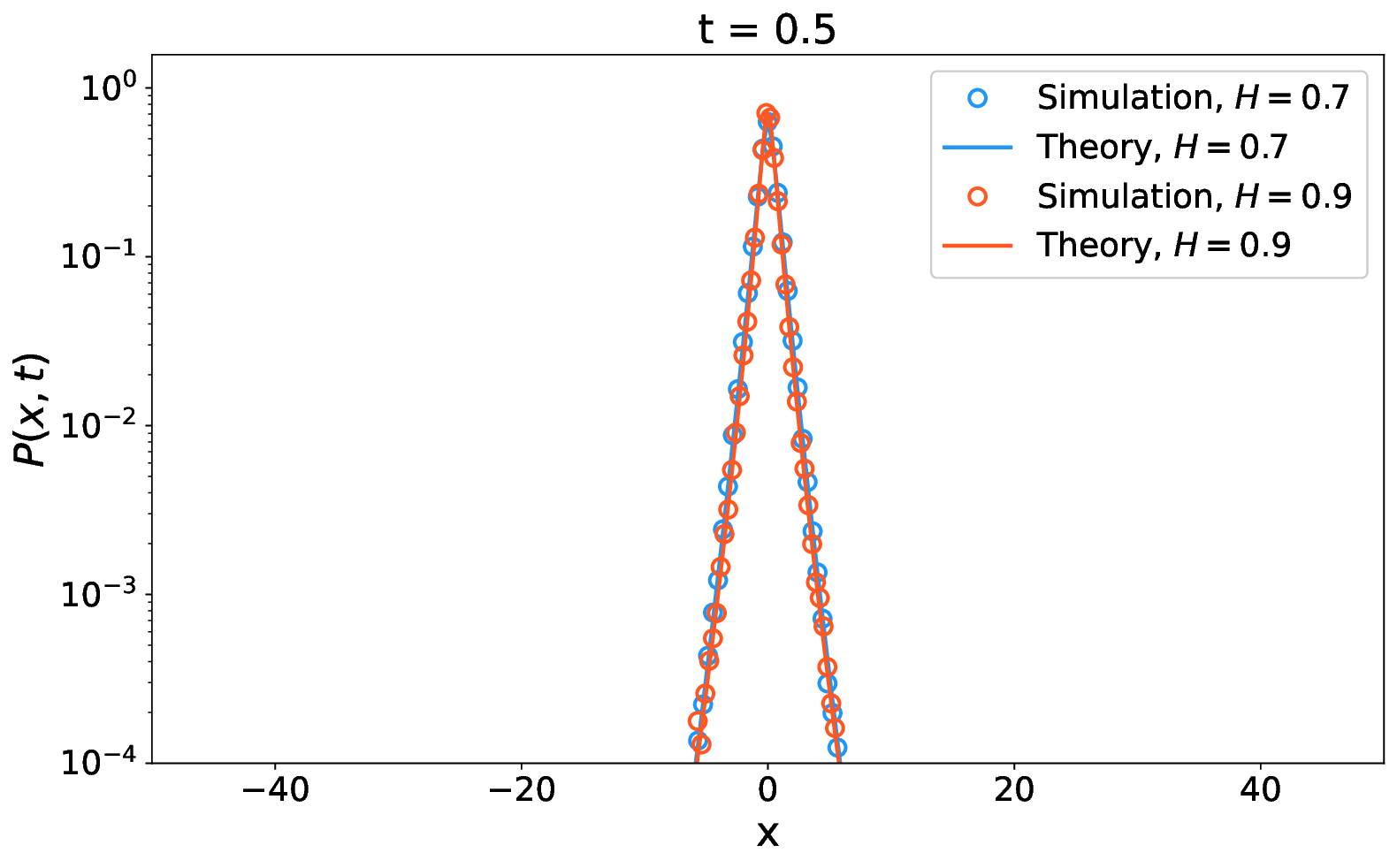}
(b)\includegraphics[width=7.2cm]{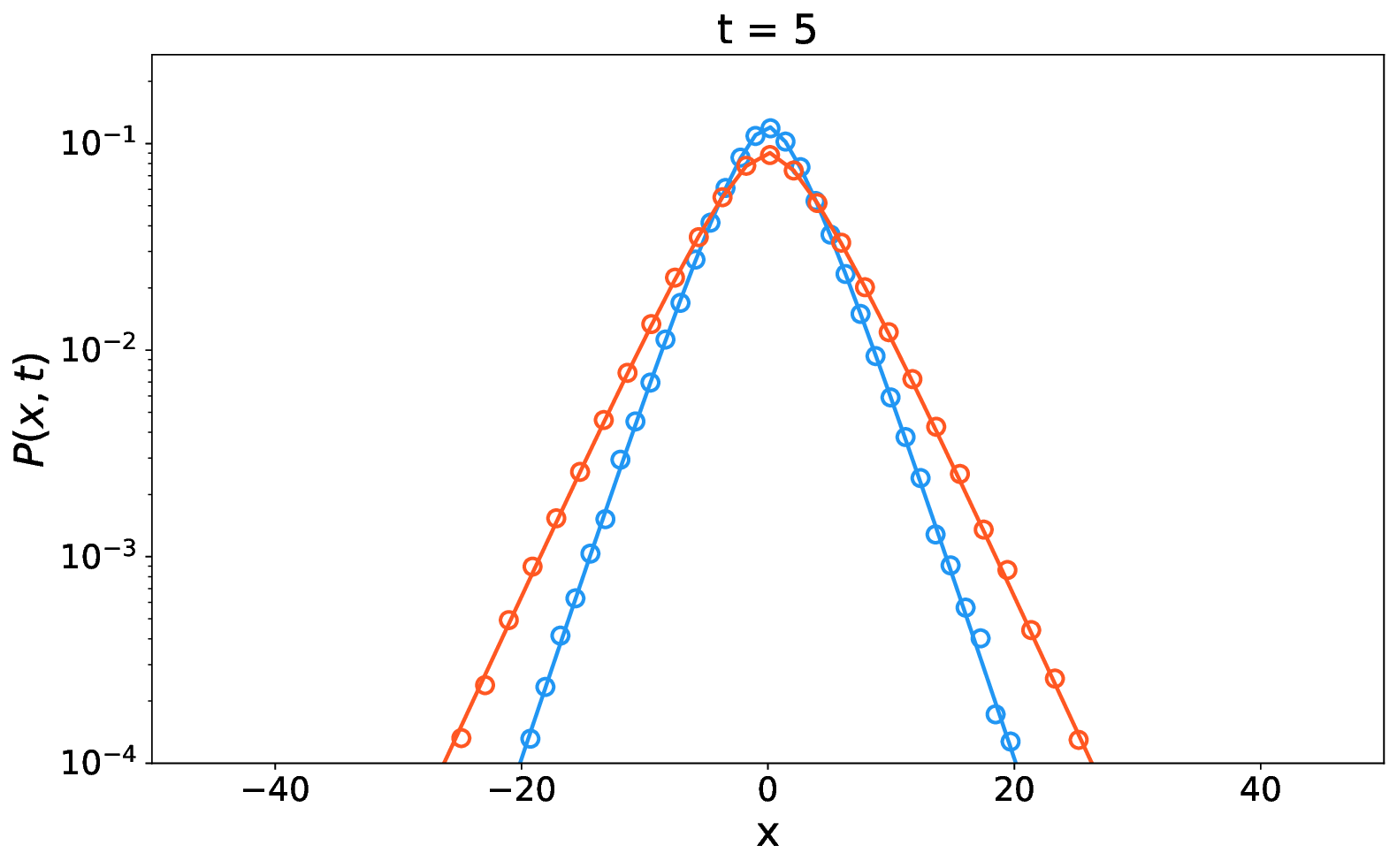}\\
(c)\includegraphics[width=7.2cm]{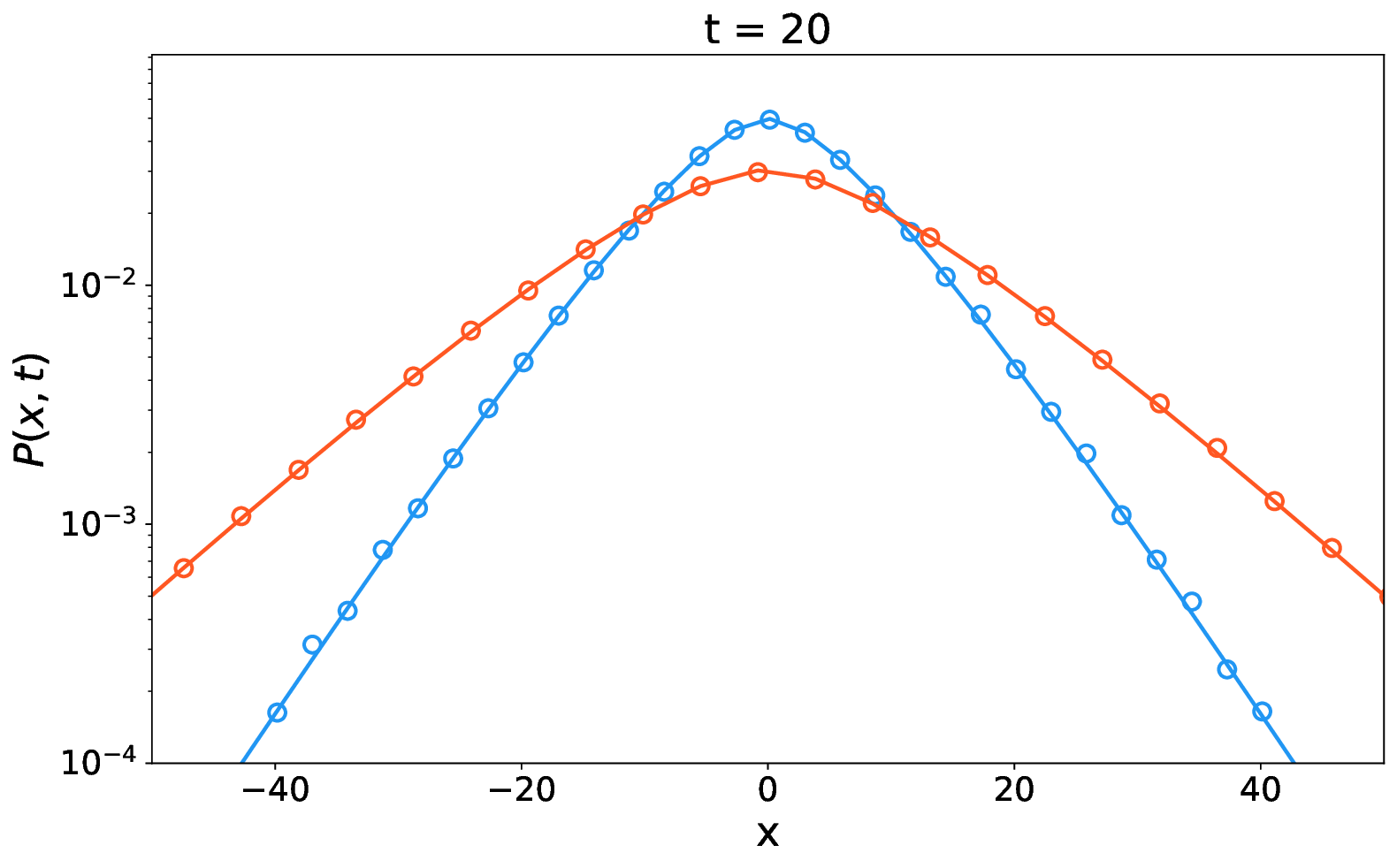}
(d)\includegraphics[width=7.2cm]{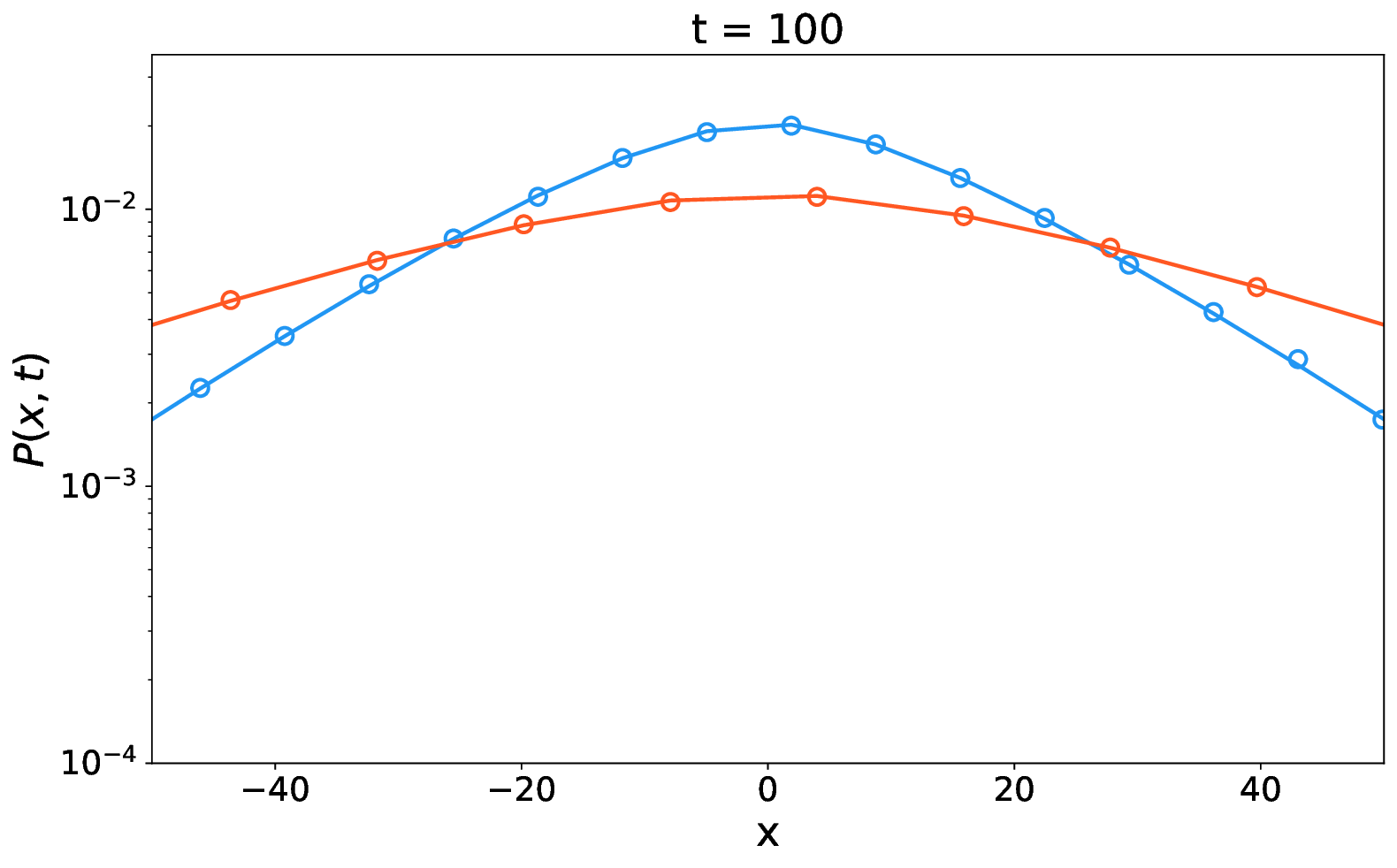}
\caption{PDF for exponentially-tempered FLM (H=0.7,0.9) with gamma-distributed diffusivity at
four different time points ($t=0.5$, $5$, $20$, and $100$). The solid coloured curves show the
theoretical prediction, via numerical integral of equation \eqref{eq-numepdf}, and the circles
represent simulations results.
\label{fig:pdf}}.
\end{figure*}

\subsection{PDF}

The displacement PDF of an ensemble of particles with different diffusivities is obtained by
averaging the propagator $G(x,s|D)$ for a specific value of the diffusivity, with respect to
the distribution of diffusion coefficients $f(D)$ \cite{chechkin2017},
\begin{equation}
\label{equ:pdf1}
P(x,t)=\int_0^{\infty}\int_0^{\infty}G(x,s|D)h(s,t)f(D)dsdD.
\end{equation}
We first perform the average over the diffusivity $D$,
\begin{eqnarray}
\nonumber
\phi(x,s)&=&\int_0^{\infty}G(x,s|D)f(D)dD\\
\nonumber
&=&\frac{1}{\sqrt{\pi}\Gamma(\alpha)\sigma(s)\sqrt{D_\beta}}\left(\frac{|x|}{2\sigma(s)\sqrt{D_
\beta}}\right)^{\alpha-\frac{1}{2}}\\
&&\times K_{\alpha-\frac{1}{2}}\left(\frac{|x|}{\sigma(s)\sqrt{D_\beta}}\right). 
\end{eqnarray}
where $K_z(x)$ is the modified Bessel function \cite{stegun}. The full PDF (\ref{equ:pdf1}) then
takes on the form
\begin{eqnarray}
\nonumber
P(x,t)&=&\frac{\lambda^{\lambda t}}{\sqrt{\pi}\Gamma(\lambda t)\Gamma(\alpha)\sqrt{D_\beta}}
\int_{0}^{\infty}s^{\lambda t-1}e^{-\lambda s}\frac{1}{\sigma(s)}\\
&&\hspace*{-1.2cm}\times\left(\frac{|x|}{2\sigma(s)\sqrt{D_\beta}}\right)^{\alpha-\frac{1}{2}}
K_{\alpha-\frac{1}{2}}\left(\frac{|x|}{\sigma(s)\sqrt{D_\beta}}\right)ds,
\label{eq-numepdf}
\end{eqnarray}
which can then be obtained by numerically evaluating the remaining integral over the operational
time. The resulting PDFs are shown in Fig.~\ref{fig:pdf}.

We further obtain the asymptotic form of the displacement PDF at long times. For $\lambda t\gg1$,
the Gamma random clock becomes self-averaging, i.e., $\langle s(t)\rangle=t$ and $\mathrm{Var}[s
(t)]=t/\lambda$, such that the relative fluctuations of the operational time decay as $(\lambda
t)^{-1/2}$. Consequently, the subordinator density $h(s,t)$ becomes increasingly concentrated
around $s=t$. For a given diffusivity $D$, the subordinated process therefore approaches the
Gaussian propagator of the parental process evaluated at $s=t$,
\begin{eqnarray}
\nonumber
P(x,t|D)&=&\int_0^\infty G(x,s|D)h(s,t)ds\\
&\sim&\frac{1}{\sqrt{4\pi D\sigma^2(t)}}\exp\left(-\frac{x^2}{4D\sigma^2(t)}\right).
\label{equ:pdf2}
\end{eqnarray}
Averaging equation \eqref{equ:pdf2} over the gamma distribution of diffusivities, we obtain the
long-time non-Gaussian PDF,
\begin{eqnarray}
\nonumber
P(x,t)&=&\int_0^\infty P(x,t|D)f(D)dD\\
\nonumber
&\sim&\frac{1}{\sqrt{\pi}\Gamma(\alpha)\sigma(t)\sqrt{D_\beta}}\left(\frac{|x|}{2\sigma(t)\sqrt{
D_\beta}}\right)^{\alpha-\frac{1}{2}}|\\
&&\times K_{\alpha-\frac{1}{2}}\left(\frac{|x|}{\sigma(t)\sqrt{D_\beta}}\right).
\label{equ:pdf_long}
\end{eqnarray}
Thus, although the fluctuations of the Gamma random clock progressively self-average, the
heterogeneity of the diffusivity distribution survives and prevents the displacement PDF
from becoming Gaussian.

For times additionally satisfying $t\gg\tau$, the variance $\sigma^2(t)$ in equation
(\ref{eq-sigma}) has the asymptotic form
\begin{equation}
\sigma^2(t)\sim\Gamma(2H+1)\tau^{2H-1}t.
\end{equation}
It is then convenient to introduce the characteristic displacement scale
\begin{eqnarray}
\ell(t)&=&\sqrt{D_\beta\Gamma(2H+1)\tau^{2H-1}t}.
\end{eqnarray}
With this expression, equation \eqref{equ:pdf_long} can be written as
\begin{equation}
\label{equ:pdf_long2}
P(x,t)\sim\frac{1}{\sqrt{\pi}\Gamma(\alpha)\ell(t)}\left(\frac{|x|}{2\ell(t)}\right)^{\alpha
-\frac{1}{2}}K_{\alpha-\frac{1}{2}}\left(\frac{|x|}{\ell(t)}\right).
\end{equation}
For large displacements, $|x|\gg\ell(t)$, with the approximation $K_\nu(z)\sim\sqrt{\pi/(2z)}e^{-z}$
of Bessel function \cite{stegun}, we further obtain
\begin{equation}
\label{equ:pdf_long_tail}
P(x,t)\sim\frac{1}{2^\alpha\Gamma(\alpha)\ell(t)}\left(\frac{|x|}{\ell(t)}\right)^{\alpha-1}
\exp\left(-\frac{|x|}{\ell(t)}\right).
\end{equation}
Hence, the long-time displacement PDF exhibits exponential tails. This demonstrates that the
recovery of the linear MSD at long times does not imply Gaussian displacement statistics: the
Gamma clock self-averages, whereas the quenched diffusivity heterogeneity remains.

\section{Conclusion}
\label{sec:Conclusion}

We studied an exponentially-tempered FLM model with distributed diffusivities to describe
complex non-Gaussian particle transport with transiently superdiffusive dynamics. Our model
combines three physical ingredients: (i) a parental FBM with exponential tempering of the
driving fractional Gaussian noise; (ii) a gamma-process subordinator, and (iii) a
gamma-distribution of diffusivities across the particle ensemble. We derived analytical
expressions for the MSD, the displacement PDF, the kurtosis, and the non-Gaussianity
parameter, all validated via stochastic simulations. We found that the model quantitatively
reproduces the three key experimental observations reported in our guiding experiment
\cite{grossmann2024non}: (i) a superdiffusion-to-normal diffusion crossover in the MSD,
controlled by the tempering timescale that reflects the tug-of-war between multiple cells;
(ii) the non-Gaussian displacement PDFs persisting across all measured timescales, arising
from the combination of gamma-subordination and distributed diffusivities; and (iii) the full
time evolution of the non-Gaussianity parameter $\Omega(\Delta)$, which is large at short lag
times due to the FLM correlations and saturates at the finite value $\Omega \to 1/\alpha$ at
long times due to the diffusivity heterogeneity.

The mechanisms identified here---the interplay of persistent, power-law correlated motion
and static disorder---are expected to be generic for a broad class of actively driven systems.
Crossovers from anomalous to normal diffusion have been reported in diverse active settings,
including bacterial suspensions \cite{wu2000particle,mino2013}, algal baths
\cite{leptos2009dynamics}, and active carpets \cite{guzman2021active}. Biological
environments are inherently heterogeneous: cell densities, metabolic activities, and local
adhesion properties all vary across space and time, so that the assumption of a distribution
of diffusivities appears quite natural. In fact, pronounced heterogeneities are already
present in single-cell studies of ensembles of amoeba \cite{beta}. Likewise, the stop-and-go
character encoded by the gamma subordinator is ubiquitous in living systems, where cargo
particles alternate between active pulling phases and passive waiting phases as cells bind and
unbind. The three ingredients of our model are therefore expected to apply, with suitable
parameter choices, to a wide range of composite active matter systems.

Several open questions remain. First, recent experiments indicate that even the motion of a
single \textit{D.~discoideum} cell exhibits a crossover from superdiffusion to normal
diffusion \cite{beta}, albeit at a longer crossover time of $6$ to $10$~min than the
multicellular case with crossover time of $\approx2$~min observed in \cite{grossmann2024non}.
This suggests a possible hierarchy of correlation timescales: the persistence of individual
cell motion sets a longer intrinsic cutoff, while the tug-of-war among multiple cells
attached to the same particle introduces an additional, shorter decorrelation mechanism.
A unified microscopic model that bridges single-cell and multicellular transport would be a
valuable extension of the present framework. Second, the physical origin of the distributed
diffusivities warrants further investigation. In the current model, $D$ is treated as a
static quenched disorder; however, on sufficiently long timescales a particle explores a
large fraction of the cellular environment, and the effective diffusivity may cross over to
a homogeneous, ensemble-averaged value. Verifying this prediction requires longer
single-particle tracking data than currently available. Third, the quantitative analysis of
measured or simulated data in terms of single-trajectory, finite-time power spectral analysis
developed for FBM \cite{power}, the mean-squared increment \cite{prx} as well as the scaling
exponent approach studies in \cite{ohad} will need to be extended to exponentially-tempered
FLM introduced here.

\begin{acknowledgments}
R.M. acknowledges financial
support from the German Science Foundation (DFG, Grant No. 318763901, CRC Data Assimilation and
ME 1535/22-1) and NSF-BMFTR CRCNS (Grant 2112862/STAXS).
\end{acknowledgments}

\section*{Data availability}

The code used to generate the simulation results and figures is available at \cite{code}.

\appendix

\section{Derivation of the MSD and its asymptotic behaviour}
\label{app-a}

Using equation (8.14.5) of Ref.~\cite{olver2010},
\begin{eqnarray}
\nonumber
&&\int_0^\infty x^{a-1}e^{-cx}\gamma(b,x)dx\\
&&=\frac{\Gamma(a+b)}{b(1+c)^{a+b}}{}_2F_1\!\left(1,a+b;1+b;\frac{1}{1+c}\right),
\end{eqnarray}
valid for $c>0$ and $a+b>0$, where ${}_2F_1$ denotes the Gaussian hypergeometric function,
together with relation (15.5.15),
\begin{eqnarray}
\nonumber
&&\frac{1}{b}{}_2F_1\left(1,a+b;1+b;\frac{1}{1+s}\right)\\
\nonumber
&&\quad-\frac{1}{1+b}{}_2F_1\left(1,a+b;2+b;\frac{1}{1+s}\right)\\
&&=\frac{1}{b(1+b)}{}_2F_1\left(2,a+b;2+b;\frac{1}{1+s}\right),
\end{eqnarray}
the MSD in equation \eqref{eq-msd-gamma} can be expressed as
\begin{eqnarray}
\nonumber
\langle x^2(t)\rangle&=&\alpha D_\beta\tau^{2H}\frac{(\lambda\tau)^{\lambda t}\Gamma(\lambda t+2H)}
{\Gamma(\lambda t)(1+\lambda\tau)^{\lambda t+2H}}\\
&&\times{}_2F_1\left(2,\lambda t+2H;2H+1;\frac{1}{1+\lambda\tau}\right).
\end{eqnarray}
Applying the Pfaff transformation \cite{stegun}
\begin{eqnarray}
\nonumber
&&{}_2F_1\left(2,\lambda t+2H;2H+1;\frac{1}{1+\lambda\tau}\right)=\left(\frac{1+\lambda\tau}{
\lambda\tau}\right)^{\lambda t+2H}\\
&&\times{}_2F_1\left(2H-1,\lambda t+2H;2H+1;-\frac{1}{\lambda\tau}\right),
\end{eqnarray}
we then obtain the final expression \eqref{eq-msd} of the MSD.

For the asymptotic behaviour, in the short time $t\ll\lambda^{-1}$, we have $\lambda t\ll1$,
and the Gamma function ratio and hypergeometric function can be approximated by the leading
terms,
\begin{equation}
\frac{\Gamma(\lambda t+2H)}{\Gamma(\lambda t)}\sim\Gamma(2H)\lambda t,
\end{equation}
and 
\begin{eqnarray}
\nonumber
&&\hspace*{-0.8cm}{}_2F_1\left(2H-1,\lambda t+2H;2H+1;-\frac{1}{\lambda\tau}\right)\\
&&\sim1-\frac{(2H-1)(\lambda t+2H)}{(2H+1)\lambda\tau}.
\end{eqnarray}
Therefore, the MSD is linear at very short times $t\ll\lambda^{-1}$, taking the form
\begin{equation}
\langle x^2(t)\rangle\sim\alpha D_\beta\Gamma(2H)\lambda^{1-2H}t.
\end{equation}
This regime originates from the strong fluctuations of the Gamma random clock at times shorter
than its characteristic time $\lambda^{-1}$.

In the intermediate regime $\lambda^{-1}\ll t\ll\tau$, we have $\lambda t\gg1$, such that
\begin{equation}
\frac{\Gamma(\lambda t+2H)}{\Gamma(\lambda t)}\sim(\lambda t)^{2H}.
\end{equation}
In this regime, the small-argument expansion of the hypergeometric function becomes
\begin{equation}
{}_2F_1\left(2H-1,\lambda t+2H;2H+1;-\frac{1}{\lambda\tau}\right)\sim1-\frac{2H-1}{2H+1}
\frac{t}{\tau},
\end{equation}
where terms of order $1/(\lambda\tau)$ are neglected as compared to $t/\tau$ for $\lambda t\gg1$.
Thus, for $t\ll\tau$, the hypergeometric function remains close to unity, and the MSD reduces to
\begin{equation}
\langle x^2(t)\rangle\sim\alpha D_\beta t^{2H}.
\end{equation}
For $H>1/2$, this corresponds to transient superdiffusion governed by the persistent correlations
of the parental FBM.

At long times $t\gg\tau$, the hypergeometric function can no longer be approximated by unity even
though $\lambda\tau\gg1$, since its second parameter grows as $\lambda t$. Using the integral
representation
\begin{eqnarray}
\nonumber
&&{}_2F_1\left(2H-1,\lambda t+2H;2H+1;-\frac{1}{\lambda\tau}\right)=2H(2H-1)\\
&&\hspace*{0.8cm}\times\int_0^1y^{2H-2}(1-y)\left(1+\frac{y}{\lambda\tau}\right)^{-(\lambda
t+2H)}dy,
\end{eqnarray}
and noting that for $t\gg\tau$ the dominant contribution comes from $y\ll1$, we use
\begin{equation}
\left(1+\frac{y}{\lambda\tau}\right)^{-(\lambda t+2H)}\sim\exp\left(-\frac{\lambda t+2H}{
\lambda\tau}y\right).
\end{equation}
The hypergeometric function then has the asymptotic form
\begin{eqnarray}
\nonumber
&&{}_2F_1\left(2H-1,\lambda t+2H;2H+1;-\frac{1}{\lambda\tau}\right)\\
\nonumber
&&\hspace*{1.2cm}\sim\Gamma(2H+1)\left(\frac{\lambda\tau}{\lambda t+2H}\right)^{2H-1}\\
&&\hspace*{1.2cm}\sim\Gamma(2H+1)\left(\frac{\tau}{t}\right)^{2H-1}.
\end{eqnarray}
At the same time,
\begin{equation}
\frac{\Gamma(\lambda t+2H)}{\Gamma(\lambda t)}\sim(\lambda t)^{2H}.
\end{equation}
Substitution into equation (\ref{eq-msd}) gives
\begin{equation}
\langle x^2(t)\rangle\sim\alpha D_\beta\Gamma(2H+1)\tau^{2H-1}t,\qquad t\gg\tau,
\end{equation}
showing the recovery of long-time normal diffusion due to the finite correlation time of the
parental process.

\section{Bound of $A(t)/B(t)$}
\label{app-b}

The lower bound can be estimated by the inequality
\begin{eqnarray}
\nonumber
\left[1+\epsilon(x+y)\right]^{q+2H}&=&\left[1+\epsilon(x+y)\right]^{q}\left[1+\epsilon
(x+y)\right]^{2H}\\
\nonumber
&&\hspace*{-0.8cm}<(1+2\epsilon)^{2H}\left[1+\epsilon(x+y)\right]^q\\
&&\hspace*{-0.8cm}<(1+2\epsilon)^{2H}(1+\epsilon y)^{q}(1+\epsilon x)^q,
\end{eqnarray}
for $x,y\in[0,1]$ and $\epsilon,q,H>0$. 

If we set $\epsilon=\nu,q=\eta t+2H$, we have
\begin{equation}
A(t)>B(t)\times\left(1+\dfrac{1}{\lambda\tau}\right)^{-2H}
\end{equation}
for $A(t),B(t)>0$, and we can get the lower bound in equation \eqref{equ:bound}.

The upper bound is given by
\begin{equation}
\label{equ:upper_bound}
\lim_{t\to\infty}\frac{A(t)}{B(t)}=1\qquad\frac{\dd}{\dd t}\frac{A(t)}{B(t)}>0,
\end{equation}
which we prove in the following. To simplify notation, let
\begin{equation}
b=2H-1>0,\qquad q=\lambda t+2H,\qquad p=q+2H.
\end{equation}
Define $(d\nu(y)=(1-y)y^{b-1}\,dy$. Then,
\begin{equation}
I(q)=\int_0^1(1+\nu y)^{-q}\,d\nu(y)
\end{equation}
and
\begin{equation}
G(q)=\iint_{[0,1]^2}(1+\nu(y+z))^{-p}\,d\nu(y)\,d\nu(z),
\end{equation}
So, we see that
\begin{equation}
\dfrac{A(t)}{B(t)}\equiv R(q)=\frac{G(q)}{I(q)^2}.
\end{equation}

First, we prove that $R(q)$ increases monotonically with $q$. Let $L(y)=\log(1+\nu y)$ and
define the normalised probability measure
\begin{equation}
\nu_q(y)dy\equiv\dfrac{ (1+\nu y)^{-q}\,d\nu(y)}{I(q)}
\end{equation}
and
\begin{equation}
Q_q(y,z)dydz=\dfrac{(1+\nu(y+z))^{-p}d\nu(y)d\nu(z)}{G(q)}.
\end{equation}
We now define the expectation with $\nu_q(y)$ and $Q_q(y,z)$,
\begin{equation}
E_{\nu_q}[L(y)]\equiv\int f(y)d\nu_q(y)dy
\end{equation}
and
\begin{equation}
E_{Q_q}[f(y,z)]\equiv\int f(y,z)Q_q(y,z)dydz.
\end{equation}
Direct differentiation produces
\begin{eqnarray}
\nonumber
\frac{d}{dq}\log R(q)&=&2E_{\nu_q}[L(Y)]-E_{Q_q}[L(Y+Z)]\\
&>&2\left[E_{\nu_q}[L(Y)]-E_{Q_q}[L(Y)].
\right]
\label{eq:log-der}
\end{eqnarray}

Let $M_q(y)$ be the $Y$-marginal density of $Q_q(y,z)$. The likelihood ratio of $M_q$ and $\nu_q$
is
\begin{equation}
\rho(y)\equiv\frac{M_q(y)}{\nu_q(y)}=\frac{I(q)}{G(q)}(1+\nu y)^q\int_0^1[1+\nu(y+z)]^{-p}d\nu(z).
\end{equation}
Set $a=1+\nu y$ , $x=\nu/a$ and $u=zx$. A change of variables gives
\begin{equation}
\rho(y)\propto J(x)\equiv\int_0^x(x-u)u^{b-1}(1+u)^{-p}du.
\end{equation}
The right-hand side has the derivative
\begin{equation}
J'(x)=\int_0^x u^{b-1}(1+u)^{-p}du>0.
\end{equation}
Thus, $J(x)$ is strictly increasing in $x$. Since $x=\nu/(1+\nu y)$ is strictly decreasing in
$y$, the density ratio $\rho(y)=M_q(y)/\nu_q(y)$ is strictly decreasing on $(0,1)$.

Recall that $M_q(y)=\int_0^1 Q_q(y,z)dz$ is the $Y$-marginal density of $Q_q$. Therefore,
\begin{equation}
E_{Q_q}[L(Y)]=E_{M_q}[L(Y)].
\end{equation}
Using $\int_0^1\rho(y)\nu_q(y)dy=1$, we have
\begin{eqnarray}
\nonumber
&&E_{M_q}[L(Y)]-E_{\nu_q}[L(Y)]\\
\nonumber
&&\hspace*{0.6cm}=\frac{1}{2}\int_0^1\int_0^1[L(y)-L(z)][\rho(y)-\rho(z)]\\
&&\hspace*{2.6cm}\times\nu_q(y)\nu_q(z)\,dy\,dz<0.
\end{eqnarray}
The strict inequality follows because $L(y)=\log(1+\nu y)$ is strictly increasing, $\rho(y)$ is
strictly decreasing, and $\nu_q(y)$ is positive on $(0,1)$. Consequently,
\begin{equation}
E_{Q_q}[L(Y)]=E_{M_q}[L(Y)]<E_{\nu_q}[L(Y)].
\end{equation}
Along with relation \eqref{eq:log-der},  we thus have
\begin{equation}
\frac{d}{dq}R(q)>0.
\end{equation}

In the long time limit, as $q\to\infty$, the integrals are dominated by $y,z\sim 0$. The standard
Laplace method gives
\begin{equation}
I(q)\sim\frac{\Gamma(b)}{(\nu q)^b},\qquad G(q)\sim\frac{\Gamma(b)^2}{(\nu q)^{2b}}.
\end{equation}
Therefore,
\begin{equation}
\lim_{q\to\infty}R(q)=\lim_{q\to\infty}\frac{G(q)}{I(q)^2}=1.
\end{equation}

Since $R(q)$ is strictly increasing in $q$, and $q=\lambda t+2H$ is strictly increasing in $t$,
we obtain
\begin{equation}
\frac{A(t)}{B(t)}<1
\end{equation}
for all finite $t$, and
\begin{equation}
\lim_{t\to\infty}\frac{A(t)}{B(t)}=1.
\end{equation}

\section{Simulations details}
\label{app:simulations}

Here we provide the details for the simulations used to generate trajectories of the tempered
FLM process with distributed diffusivities. All results are computed in one dimension; the
generalisation to higher dimensions is straightforward by treating each spatial coordinate as
an independent copy of the one-dimensional process.

The simulation for FLM proceeds in three steps: (i) generation of the parental FBM trajectory,
(ii) generation of the gamma subordinator, and (iii) interpolation at subordinated times. For a
given value of the generalised diffusion coefficient $D$, the FBM trajectory $B_H(s)$ is
generated at discrete operational times $s_k=k\times ds$ with the internal time step $ds=dt/2$,
where $dt=0.1$ is the physical time step of the composite process. FBM is obtained via the
Davies-Harte circulant embedding method \cite{dieker2004simulation}. Simulations were performed
for Hurst exponents $H=0.7$ and $H=0.9$, both with the exponential cutoff timescale $\tau=10$.

Independently, the gamma subordinator $s(t_n)$ at discrete laboratory times $t_n=n\times dt$
is generated by sampling gamma-distributed increments with shape parameter $\eta dt$ and
scale parameter $1/\lambda$, and cumulatively summing them. We set $\eta=\lambda=10$, so that
the characteristic timescale of the gamma process is unity and $\langle s(t)\rangle=t$. The
composite trajectory $x(t_n)=B_H(s(t_n))$ is then obtained by linear interpolation of the FBM
trajectory onto the subordinator times,
\begin{equation}
x(t_n)=B_H(s_{k_n})+\frac{s(t_n)-s_{k_n}}{h}\left[B_H(s_{k_n+1})-B_H(s_{k_n})\right],
\end{equation}
where $k_n=\lfloor s(t_n)/ds\rfloor$ is the integer obtained by use of the floor function
$\lfloor\cdot\rfloor$.

To incorporate distributed diffusivities, we draw $N_D=5000$ independent values of $D$ from
the gamma distribution \eqref{equ:f} with shape parameter $\alpha=2$ and scale parameter
$D_\beta=1$. For each value of $D$, $N_{\mathrm{each}}=40$ independent FLM trajectories are
generated, yielding an ensemble of $N=N_D\times N_{\rm each}=200\,000$ trajectories. The
total measurement time is $T=1000$. The parameter values chosen for the simulations are
listed in table \ref{tab:simparams}.

\end{document}